\documentclass[a4paper, prb, twocolumn, showpacs]{revtex4}

\usepackage{amssymb}
\usepackage{amsmath}
\usepackage{color}
\usepackage{graphics, graphicx}
\usepackage{psfrag}
\usepackage{subfigure}

\usepackage{color}
\usepackage{soul}

\begin{document}

\author{S. Backes$^1$}
\author{I. Titvinidze$^{1,2}$}
\author{A. Privitera$^{1,3}$}
\author{W. Hofstetter$^1$}
\affiliation{1. Institut f\"ur Theoretische Physik, Johann Wolfgang Goethe-Universit\"at, 60438 Frankfurt am Main, Germany\\
2. I. Instit\"ut f\"ur Theoretische Physik, Universit\"at Hamburg, Jungiusstra\ss e 9, 20355 Hamburg, Germany\\
3. Democritos National Simulation Center, Consiglio Nazionale delle Ricerche, Istituto Officina dei Materiali (CNR-IOM) 
and International School for Advanced Studies (SISSA), I-34136 Trieste, Italy} 

\title{Monte Carlo study of fermionic trions in a square lattice with harmonic confinement}

\begin{abstract}
We investigate the strong-coupling limit of a three-component Fermi mixture in an optical lattice with attractive
interactions. In this limit bound states (trions) of the three components are formed. We study the effective Hamiltonian 
for these composite fermions and show that it is asymptotically equivalent at the leading-order to an antiferromagnetic Ising model. By
using Monte-Carlo simulations, we investigate the spatial arrangement of the trions in this regime and the formation of a
trionic density wave (CDW), both in a homogeneous lattice and in the presence of an additional harmonic confinement. Depending on
the strength of the confinement and on the temperature, we found several scenarios for the trionic distribution, including
coexistence of disordered trions with CDW and band insulator phases. Our results show that, due to a proximity effect,
staggered density modulations are induced in regions of the trap where they would not otherwise be present according to the local density
approximation. 
\end{abstract}

\maketitle

\section{INTRODUCTION}

Ultracold atomic gases trapped in optical lattices provide us with a new laboratory to investigate quantum many-body systems. 
They allow us not only to realize model Hamiltonians for electronic systems \cite{Hubbard1, Hubbard2} but also to investigate systems
which have no direct counterpart in condensed matter physics.

A shining example in this context is provided by three-component Fermi mixtures loaded into optical lattices, which have been 
theoretically investigated using a variety of approaches \cite{Honerkamp-Hofstetter, Rapp-Hofstetter, Kantian, Molina, Inaba,
Trions_short, NJP, Jan, Blumer_3flavor, Inaba2}. The results support the overall idea that these mixtures give access to very different
phenomena depending on the parameter regime under investigation, ranging from Mott insulating behavior for repulsive  interactions
\cite{Blumer_3flavor,Inaba2} to color-superfluidity and trionic phases for attractive interactions \cite{Honerkamp-Hofstetter,
Rapp-Hofstetter, Kantian, Molina, Inaba, Trions_short, NJP, Jan}. Moreover, an interesting interplay between superfluidity and
magnetism has been found to induce domain formation in globally balanced mixtures with $SU(3)$ attractive interaction \cite{NJP},
in striking contrast with the balanced two-component case. This tendency towards phase separation is quite general 
in the attractive case, being also present for asymmetric interactions in the strong loss regime \cite{Trions_short}. In addition, it has been also
pointed out \cite{Wilczek} that multi-component Fermi mixtures can help to shed some light onto very complex phenomena connected to
Quantum Chromo Dynamics.          

Testing these theoretical predictions in a laboratory is within the experimental capabilities of today, although current experiments are 
still performed without optical lattices. Indeed, mixtures of three different magnetic sublevels of $^6{\rm
Li}$~[\onlinecite{Li_exp}] or $^{173}{\rm Yb}$~[\onlinecite{Yb_exp}], as well as $^{40}{\rm K}$ or $^{87}{\rm Sr}$~[\onlinecite{regal,stellmer}], and mixtures of the two lowest magnetic states of 
$^6{\rm Li}$ with the lowest hyperfine state of $^{40}{\rm K}$~[\onlinecite{LiK_exp}], already have been successfully trapped and
cooled in current experiments.

In this paper we focus on attractive interactions for the three-component mixture and more specifically
on the strong-coupling limit, where the potential energy contribution is dominant. There is substantial evidence\cite{Rapp-Hofstetter,Inaba,NJP}
that for strong-enough attraction the system undergoes a phase transition from a color-superfluid phase, where superfluid pairs
coexist with unpaired fermions, to a so-called trionic phase, where three fermions from different components are bound together in
new fermionic particles called trions. The formation of these three-body bound states poses new questions about their spatial
arrangement and the formation of new phases involving trions as elementary objects. Several theoretical results
\cite{Molina, Honerkamp-Hofstetter, NJP} suggest that trions tend to spontaneously break the translational invariance of the
lattice into two inequivalent sublattices and give rise to staggered density modulations for a suitable range of parameters
depending on density, temperature, and dimensionality. Despite the fact that ultracold gases are charge neutral, we use 
the expression charge density wave (CDW) throughout the paper to identify this phase in analogy with the terminology used for electronic systems.

The existence of a trionic CDW phase in the presence of harmonic confinement has been investigated using density matrix
renormalization group\cite{Molina} in $D=1$, while results for the homogeneous two-dimensional case obtained within
one-loop renormalization group and mean-field approaches \cite{Honerkamp-Hofstetter} suggest that the CDW could be the dominant 
instability at half-filling. By using dynamical mean-field theory on the Bethe lattice in $D=\infty$, it was shown
that the superfluid phase is stable at half-filling against CDW in a small but finite region of coupling for weak attraction \cite{NJP}.
This suggests that the stability of the trionic CDW phase shows a marked dependence on the dimensionality. 

In the present work we consider a two-dimensional square lattice for very strong attraction both in the homogeneous case and also with
harmonic confinement. In order to restrict the number of parameters,
 we focus on a globally balanced system, where $N_\sigma=N$ and $N_\sigma$ is the total number of particles for the component
$\sigma$. We anticipate that our results show the existence of various possible scenarios for the spatial arrangement of trions
once trapped and loaded into an optical lattice depending on the strength of the harmonic confinement and temperature.

The paper is organized as follows: in the next section we introduce a model Hamiltonian to describe a three-component mixture and 
we examine the effective trionic Hamiltonian derived in Ref. [\onlinecite{NJP}] as a strong-coupling limit of this model. 
We show that this effective Hamiltonian can be mapped asymptotically at the leading order onto an antiferromagnetic Ising model 
which we address using Monte Carlo techniques. Details on the Monte-Carlo technique used are provided in Sec. \ref{method}, while 
the results concerning both the homogeneous case and in presence of the harmonic confinement are given in Sec. \ref{results}. Finally, Sec. \ref{conclusion} concludes with a summary of the salient points of this paper.

\section{Model}
\label{sec:model}
A three-component mixture of fermions loaded into an optical lattice can be suitably described by the following single-band Hubbard Hamiltonian:
\begin{eqnarray}
\label{Initial_Hamiltonian}
{\cal H}=&-&J \sum_{\langle i,j \rangle,\sigma} c_{i,\sigma}^\dagger c_{j,\sigma}^{\phantom\dagger}+\sum_{i,\sigma>\sigma'} U_{\sigma\sigma'}
n_{i,\sigma}n_{i,\sigma'} \nonumber \\
&-&\sum_{i,\sigma} (\mu_\sigma-\frac{V_0}{3r_p^2}r_i^2) n_{i,\sigma} \, .
\end{eqnarray}
Here $c_{i,\sigma}\ (c_{i,\sigma}^\dagger)$ is the annihilation (creation) operator of fermions with hyperfine state $\sigma$ ($\sigma=1,2,3$) on the lattice site $i$ and
$n_{i,\sigma}= c_{i,\sigma}^\dagger c_{i,\sigma}^{\phantom\dagger}$ is the fermionic number operator. $J$ is the hopping amplitude between nearest neighboring
sites $\langle i,j \rangle$, $U_{\sigma\sigma'}$ the on-site two particle interaction between fermions in different 
hyperfine states $\sigma$ and $\sigma'$ and $\mu_\sigma$ is the chemical potential for the species $\sigma$. 
The harmonic confining potential is introduced using the maximally packed radius $r_p=a\sqrt{N/\pi}$ as in Ref. [\onlinecite{bloch}],
where
$a$ is the lattice spacing, and its strength is parametrized by $V_0$. The experimental constraints for realizing this model by loading a three-component Fermi mixture 
(e.g.  $^6{\rm Li}$~[\onlinecite{Li_exp}] or $^{173}{\rm Yb}$~[\onlinecite{Yb_exp}]) into an optical lattice have been already discussed in Refs. [\onlinecite{NJP, Trions_short}].  

As outlined in the introduction, this model exhibits a rich variety of physical phenomena. Here, however, we focus on the strong-coupling trionic phase where bound states of 
the three different components are formed \cite{NJP, Inaba, Rapp-Hofstetter} and we can directly describe the system in terms of the composite trions. The formation of these 
composite particles, i.e. the trionic transition, has been extensively discussed, together with their stability against three-body losses, in Refs.  [\onlinecite{NJP,Trions_short, Jan}]. 
Here we point out that the strong-coupling regime, i.e. small $J/|U_{\sigma\sigma'}|$, which is the  main point of interest in this work, can be realized in two different ways: (i) by tuning the magnetic 
field and correspondingly changing the scattering length, i.e. increasing $|U_{\sigma\sigma'}|$, and (ii) by increasing the depth of the optical lattice (decreasing $J$).  Due to the strong dependence 
of the three-body loss rate $\gamma_3$ on the applied magnetic field, at least in the case of $^6{\rm Li}$ [\onlinecite{Li_exp}], the results presented in the  manuscript essentially apply to the case of 
cold gases whenever three-body losses are negligible, that is $\gamma_3 \ll 1$, and  $ J \ll |U_{\sigma\sigma'}|$.

As we showed in Ref. [\onlinecite{NJP}], an effective Hamiltonian for the trions can be derived by applying strong-coupling perturbation theory  ($J/|U_{\sigma\sigma^\prime}| \ll 1$) to the original
Hamiltonian in Eq.~(\ref{Initial_Hamiltonian}). By keeping only the leading-order and next-to-leading-order terms, which correspond to second- and third-order virtual hopping processes of the original 
fermions, the Hamiltonian has the following form  
\begin{eqnarray}
\label{Trionic_Hamiltonian}
{\cal H}_{\rm eff}&=&-J_{\rm eff} \sum_{\langle i,j\rangle}t_i^\dagger t_j + V_{\rm eff}\sum_{\langle i,j\rangle} n_i^T n_j^T  \nonumber \\
&&- \sum_i (\mu_{\rm eff}-\frac{V_0}{r_p^2}r_i^2)n_i^T\, .
\end{eqnarray}
where $t_i\ (t_i^\dagger)$ is the annihilation (creation) operator of a local trion at lattice site $i$ and $n_{i}^T =t_i^\dagger t_i$ is the trionic number operator. Since trions are color singlets, 
this Hamiltonian is analogous to a spinless fermion model with nearest-neighbor interaction in the presence of an external field. As a consequence of the Pauli principle double occupancies for 
trions are forbidden. The effective trionic hopping parameter $J_{\rm eff}$, the effective interaction between trions in the nearest neighboring sites $V_{\rm eff}$ and the chemical 
potential $\mu_{\rm eff}$ are given respectively by
\begin{subequations}
\begin{eqnarray}
&&\hspace{-1cm}J_{\rm eff}=\sum_{\sigma,\sigma^\prime}\frac{J^3}{(U_{\sigma,\sigma'}+U_{\sigma,\sigma''}) (U_{\sigma,\sigma''}+U_{\sigma',\sigma''})}
\label{Jeff_0} \\
&&\hspace{-1cm}V_{\rm eff}= -\sum_{\sigma}\frac{J^2}{U_{\sigma,\sigma'}+U_{\sigma,\sigma''}} \label{Veff_0}  \\
&&\hspace{-1cm}\mu_{\rm eff}=\sum_\sigma \mu_\sigma+\sum_{\sigma>\sigma'}U_{\sigma,\sigma'}+ \sum_{\sigma}\frac{zJ^2}{U_{\sigma,\sigma'}+U_{\sigma,\sigma''}}
\label{mueff_0},
\end{eqnarray}
\end{subequations}
where in the sum $\sigma$, $\sigma'$ and $\sigma''$ are different from each other and $z$ is the number of the nearest neighbors 
\cite{NJP}. For the $SU(3)$-symmetric case these expressions simplify \cite{NJP, csaba_unpublished,klingshat} to
\begin{subequations}
\begin{align}
\label{Jeff}
J_{\rm eff}&=\frac{3J^3}{2U^2} && \\
\label{Veff}
V_{\rm eff}&= \frac{3J^2}{2|U|} && \\
\mu_{\rm eff}&=3(\mu_\sigma+U)-\frac{3zJ^2}{2|U|}, &&
\end{align}
\end{subequations}
where we assumed $U < 0$ for attractive interactions.

From  Eqs~(\ref{Jeff}) and (\ref{Veff}), it follows that, due to the specific features of the underlying three-component Fermi mixture \cite{virtual},  $J_{\mathrm{eff}}/V_{\mathrm{eff}}=J/|U| \ll 1$ and therefore the effective trionic Hamiltonian intrinsically map onto the strong-coupling limit of a spinless 
fermion model, while the corresponding weak-coupling regime does not have a counterpart in terms of the three-component mixture. Indeed the trions are not yet formed 
for small values of the interaction and the system is in a color-superfluid state \cite{NJP}. 

The effective Hamiltonian is asymptotically dominated by the interaction term, while the kinetic term is only a subleading contribution in the strong-coupling regime.  Incidentally, 
it is worth pointing out that the analogous strong-coupling approach to a two-component mixture leads instead to a {\it{bosonic}} model, whose kinetic energy is of the same 
order as the interaction term \cite{capone_dao}. 

At the leading order the trionic hopping $J_{\rm eff}$ is identically zero and the effective Hamiltonian reduces to
 \begin{equation}
\label{Hamiltonian}
{\cal H}_{\rm eff} = V_{\rm eff}\sum_{\langle i,j\rangle} n_i^T n_j^T  - \sum_i (\mu_{\rm eff}-V_i)n_i^T ,
\end{equation}
where $V_i=V_0 (r_i/r_p)^2$. It is easy to realize that the Hamiltonian in Eq.~(\ref{Hamiltonian}) has a structure very similar to
the one of an antiferromagnetic Ising model in a magnetic field $B_i$
\begin{equation}
\label{Ising_Hamiltonian}
{\cal H}_{\rm Ising}=I \sum_{\langle i,j\rangle}s_i s_j -\sum_i B_i s_i,
\end{equation}
where the parameters of the two Hamiltonians are related by
\begin{subequations}
\begin{eqnarray}
\label{mapping_Veff}
&&I=\frac{1}{4}V_{\rm eff} \\
\label{mapping_mueff} 
&&B_i=\frac{1}{2}\left(\mu_{\rm eff} -V_i\right)-V_{\rm eff} \\
\label{operators}
&&s_i=2n_i^T-1 .
\end{eqnarray}
\end{subequations}
with the correspondence  "trion"$=\uparrow$ and "no trion"$=\downarrow$.
Correspondingly, the homogeneous system ($V_0=0$) directly maps onto an antiferromagnetic Ising model in the presence of a uniform magnetic field, while the effect of the trap is equivalent to a non-uniform magnetic field profile. A similar mapping on a spin model has also been used to investigate two-component mixtures with hopping imbalance \cite{capone_dao}.

\section{METHOD} \label{method}

In order to investigate the spatial arrangement of trions we perform Monte Carlo simulations 
of a two-dimensional lattice with $M$ sites. By exploiting the mapping established 
in the previous section the results can be related to the equivalent Ising spin model.  

The probability in the grand canonical ensemble for a specific configuration $\{n_i^T \}$ with temperature $T=1/k_B\beta$ is given by 
\begin{align}
p&=Z^{-1}\mathrm{e}^{-\beta E(\{n_i^T \})} \nonumber \\
&=Z^{-1}\exp\left\{-\beta \left( V_{\rm eff}\sum_{\langle i,j\rangle} n_i^T n_j^T  - \sum_i (\mu_{\rm eff}-V_i)n_i^T \right)  \right\}.
\label{eq:probability}
\end{align}
The configuration space $\{n_i^T \}$ was sampled by using a Markov-chain approach \cite{Krauth, Newman} which is based on
creation or annihilation of a single trion in a given lattice site. This corresponds to a single spin flip in the equivalent
spin model.
In order to have an efficient strategy for the configuration updates, we used the Metropolis algorithm \cite{Metropolis}, where   
the transition probability $\bar{p}(a\rightarrow b)$ between two configurations is given by 
\begin{equation}
\bar{p}(a\rightarrow b)=min\left[ 1,\frac{p(b)}{p(a)} \right],
\label{eq:metropolisalgorithm}
\end{equation}
and the stationary probabilities $p(a),p(b)$ for given configurations $a$ and $b$ are
provided by Eq.~\eqref{eq:probability}. This allows the correct transition 
probabilities to be generated by computing only the energy differences between configuration $a$ and
$b$, connected by a single spin flip in the equivalent spin model.

For the homogeneous system, which is translationally invariant, we have used periodic boundary conditions.
In contrast, a system with harmonic confinement is not translationally invariant and periodic boundary conditions are unphysical.
Therefore, in order to address trions in a harmonic confinement, we used open boundary conditions, i.e. we set the occupation of the 
``missing" neighboring sites at the edges of the system equal to zero. 

Since the Hamiltonian Eq.~\eqref{Hamiltonian} in the absence of harmonic confinement is equivalent to an 
antiferromagnetic Ising model in a uniform magnetic field, we have used known results \cite{Monroe, Wang-Klim, Onsager}
to benchmark our simulations. We perform a careful finite-size scaling in order to confirm that the thermodynamic regime
has effectively been reached. All the results presented here are obtained from simulations performed 
on a $100 \times 100$ lattice. Moreover, in order to avoid autocorrelations in the Markov-chain sampling, 
 we also perform a careful study of the equilibration time $\tau$ and of the number of measures
 we use to compute each observable. The excellent agreement of our simulation with previous results for the homogeneous
system (see e.g. Fig. \ref{PD_without}) demonstrates the robustness of our simulations. Special care in the choice of the equilibration time 
was required to avoid unphysical results in the non-homogeneous case.  

In order to quantitatively characterize the system, we evaluated several observables such as the local $\langle n_i^T \rangle$ and    
global average occupation 
\begin{equation}
\langle n^T \rangle = \frac{1}{M} \sum_i \langle n^T_i \rangle,
\end{equation}
and the global CDW order parameter
\begin{equation}
C = \frac{1}{M} \sum_i(-1)^i \langle n^T_i \rangle.
\end{equation}

Despite the fact that a nonzero value of $C$ allows us to determine the existence of CDW order in our system, in the inhomogeneous 
case we cannot use this information to localize the regions where CDW order takes place. In this case we identify these regions by
directly looking at the density profile $\langle n^T_i \rangle$. Moreover, we further characterize the system by evaluating 
the connected density-density correlation function $\Delta(i,j)$, defined as
\begin{equation}
\Delta(i,j)=\langle (n^T_i-\langle n^T_i \rangle)(n^T_j-\langle n^T_j \rangle) \rangle.
\end{equation}
which provides us with useful information on the density fluctuations and the correlation length.

\section{RESULTS}\label{results}
\subsection{Homogeneous Case}
\label{subsec:homogeneus}
\begin{figure}[hbpt]
\includegraphics[width=0.45\textwidth]{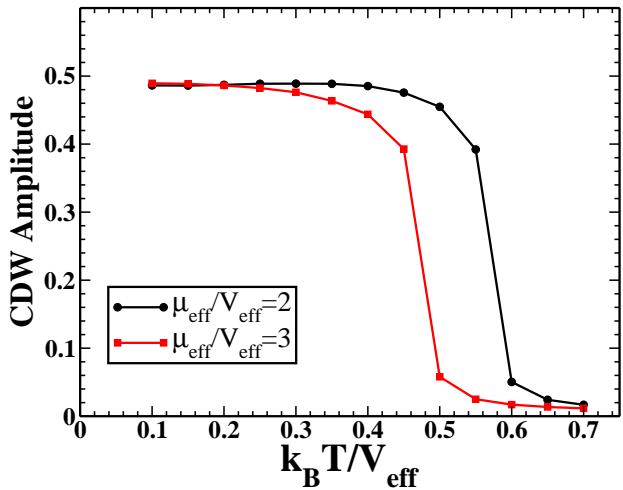}  
\vspace{-0.2cm}
\caption{(Color online) 
CDW amplitude $C$ for the homogeneous trionic system as a function of the temperature $T$ 
 for different values of the chemical potential $\mu_{\rm eff}$.}
\label{CDW_without}
\end{figure}

We first consider a homogeneous lattice system without harmonic confinement. In this case, as we mentioned above, the Hamiltonian 
(Eq.~\eqref{Hamiltonian}) can be mapped onto an antiferromagnetic Ising model (Eq.~\eqref{Ising_Hamiltonian}) in the presence of a uniform magnetic field.
Correspondingly, the results for the trionic and the Ising model can be mapped onto each other by 
the simple transformations (Eqs.~(\ref{mapping_Veff})-(\ref{operators})) and the antiferromagnetic phase in the Ising model
corresponds to the CDW phase for the trionic system. 

\begin{figure}[hbpt]
\includegraphics[width=0.45\textwidth]{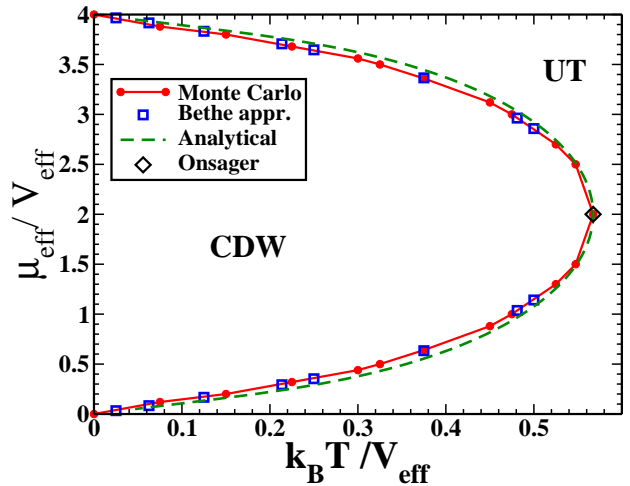}
\vspace{-0.2cm}
\caption{(Color online) Monte-Carlo phase diagram for the two-dimensional homogeneous trionic system. The red line with circles
marks the critical temperature $T_c$ for the transition from CDW ordered trions to unordered trions (UT) within our 
Monte-Carlo approach. For comparison we also plot, after a suitable rescaling given by Eqs.~(\ref{mapping_Veff})-(\ref{operators}), 
the corresponding results for the antiferromagnetic Ising model from Ref.~[\onlinecite{Monroe}] (blue squares),
 Ref.~[\onlinecite{Wang-Klim}] (dashed green line), and the Onsager solution \cite{Onsager} (black diamond).}
\label{PD_without}
\end{figure}

In order to investigate the CDW order, we first study the CDW amplitude $C$ as a function of temperature $T$ for different values of
the chemical potential, as shown in Fig. \ref{CDW_without}. At $T=0$ we found the ground state of the system to exhibit  staggered 
CDW order for $0<\mu_{\rm eff} < 4 V_{\rm eff}$ since $C \not=0$ (see Fig. \ref{PD_without}). In the homogeneous case the CDW phase
is always characterized by a commensurate density $\langle n^T \rangle=0.5$, i.e. the density modulations in the two sublattices 
are always symmetric with respect to half-filling. Outside this range of chemical potentials, that is for  $ \mu_{\rm eff} < 0$ 
or $ \mu_{\rm eff} > 4 V_{\rm eff}$, the CDW order disappears and the ground state is trivially empty $\langle n^T \rangle=0$ 
or full $\langle n^T \rangle=1$, i.e. in a band insulator phase. For $\mu=0$ and $\mu=4 V_{\rm eff}$ the CDW is degenerate 
with an unordered trionic phase, whose average density is incommensurate and not fixed by the value of the chemical potential, being 
$0< \langle n_T \rangle < 0.5$ for $\mu=0$ and $0.5 < \langle n_T \rangle < 1$ for $\mu=4 V_{\rm eff}$ respectively. 
This unordered phase has therefore an infinite compressibility and a very large degeneracy in the ground state at fixed density.
Both of these peculiar features clearly stem from having neglected the subleading kinetic energy contribution,
which would restore a finite compressibility and remove the ground state degeneracy, thus leading to a metallic phase of trions 
\cite{hoping}. Indeed, the next-to-leading-order Hamiltonian would coincide with the strong-coupling regime of the homogeneous spinless 
fermion model, e.g. studied in Refs. [\onlinecite{gubernatis,vlaming}], which shows a rich phase diagram including homogeneous metallic phases, 
commensurate and incommensurate CDW, and phase separation among these two phases.

With increasing temperature the CDW amplitude decreases and vanishes at a critical temperature $T_c$. The rounding of the transition 
and the non-zero magnetization beyond the critical temperature in Fig. \ref{CDW_without} are due to finite-size effects. We summarize
our results in the phase diagram in Fig. \ref{PD_without}. For comparison we also show results for the Ising model from Ref.~\onlinecite{Monroe}, obtained by the use of an extended Bethe approximation,
and those from Ref.~\onlinecite{Wang-Klim}, where an analytical method based on the relation between the singularities of the free energy 
and the zeros of the Ising pseudo-partition function on an elementary circle were used. As one can see, we find very good
agreement with our Monte Carlo simulations. We also compared our results with the Onsager solution \cite{Onsager} in the absence of a magnetic field and again found very good agreement. 

\begin{figure}[b]
\includegraphics[width=0.45\textwidth]{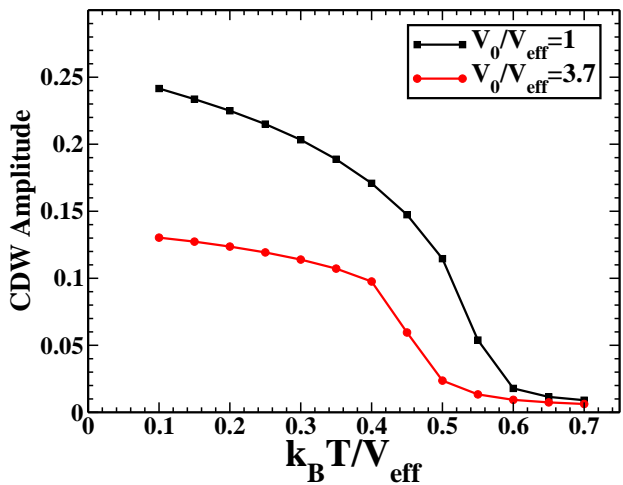}  
\vspace{-0.2cm}
\caption{(Color online) Global CDW amplitude $C$ of the system in the presence of harmonic confinement 
as a function of temperature $T$ for different values of trapping potential.}
\label{CDW}
\end{figure}

\begin{figure*}[hbpt]
\begin{center}
\subfigure[]{
\label{CDW_center}
\begin{minipage}[b]{0.45\textwidth}
\centering \includegraphics[trim = 15mm 17mm 17mm 15mm, clip, width=1\textwidth]{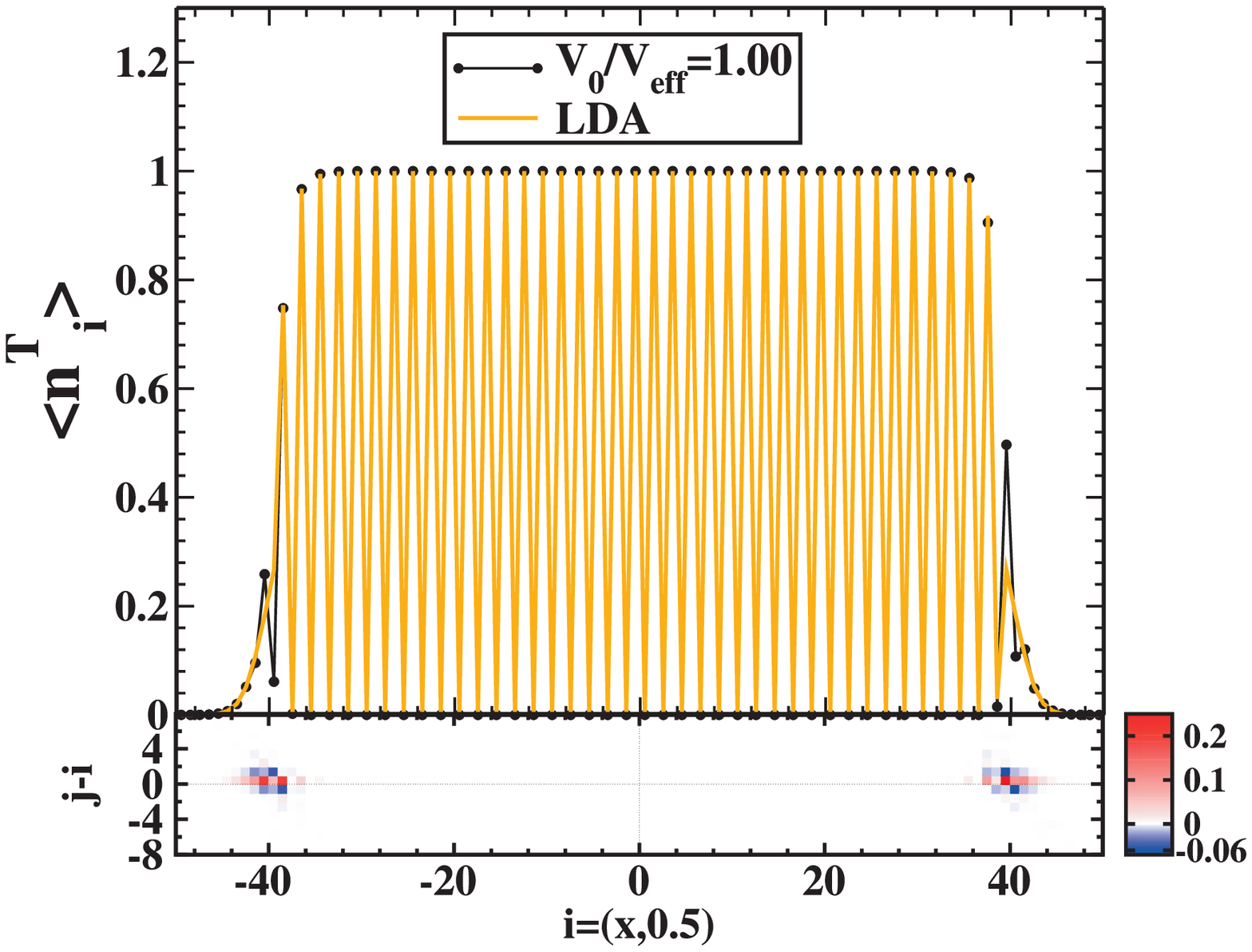}
\end{minipage}}
\hspace{0.5cm}
\subfigure[]{
\label{CDW_left}
\begin{minipage}[b]{0.45\textwidth}
\centering \includegraphics[trim = 15mm 17mm 17mm 15mm, clip, width=1\textwidth]{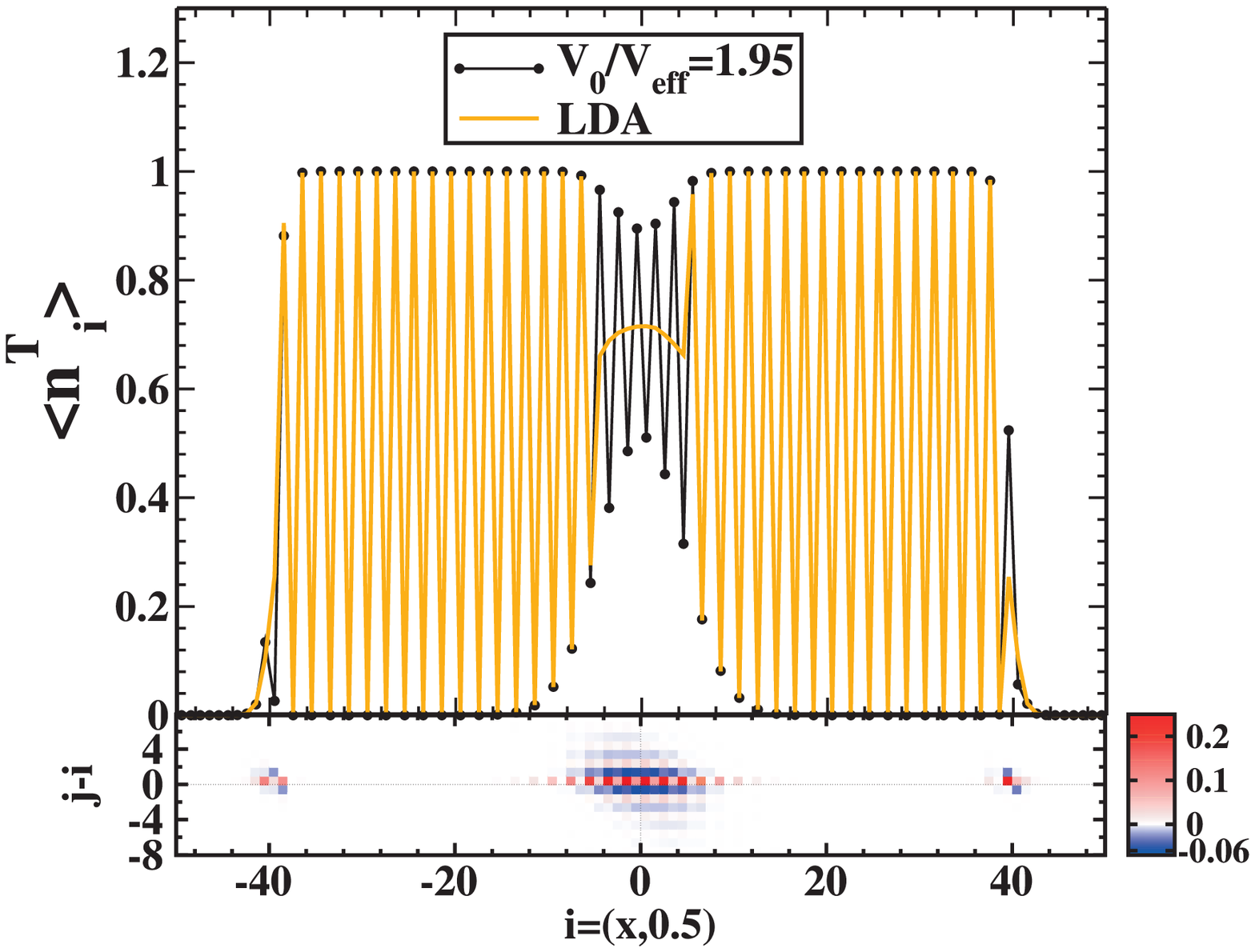} 
\end{minipage}}\\
\subfigure[]{
\label{CDW_right}
\begin{minipage}[b]{0.45\textwidth}
\centering \includegraphics[trim = 15mm 17mm 17mm 15mm, clip, width=1\textwidth]{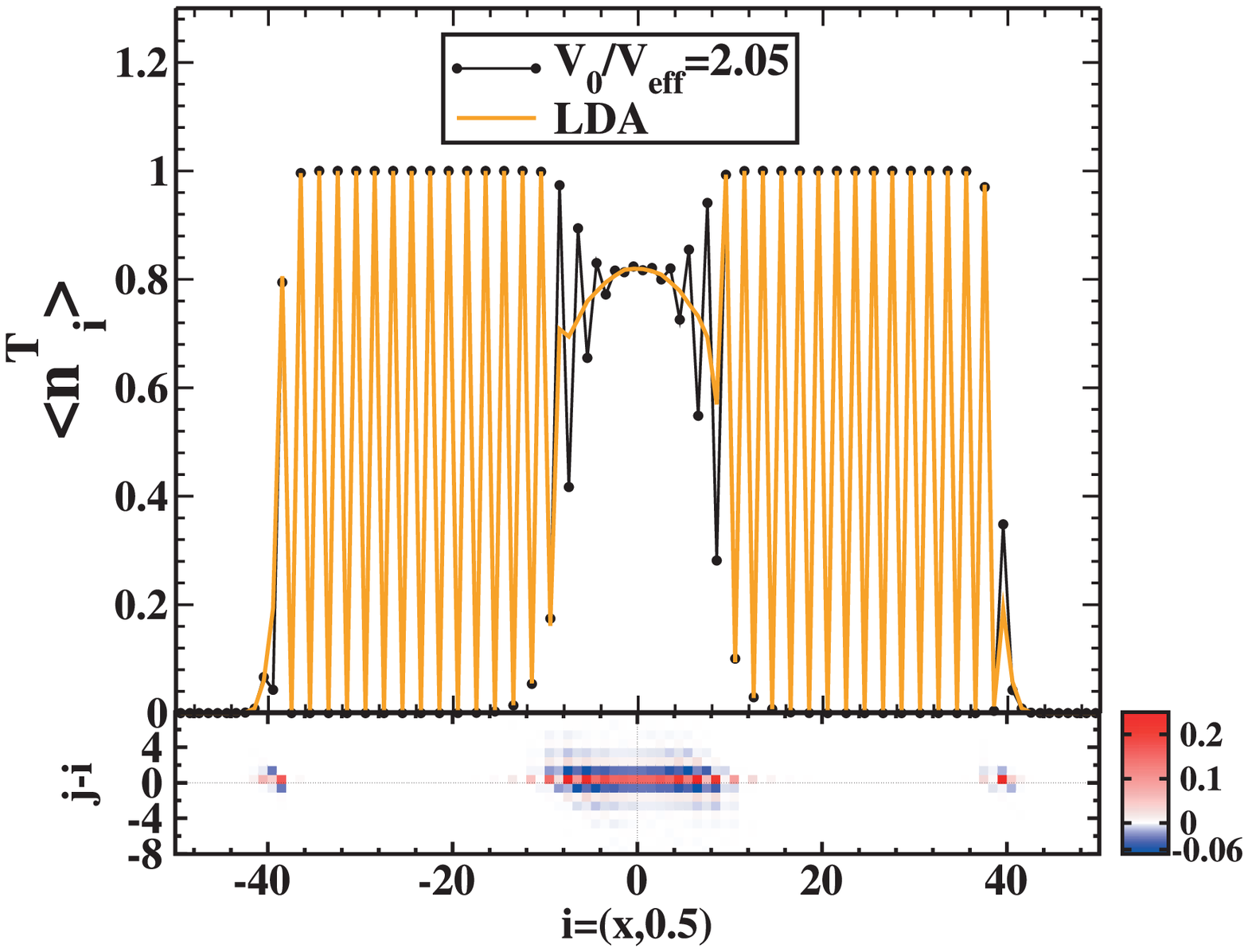}
\end{minipage}}
\hspace{0.5cm}
\subfigure[]{
\label{CDW_ring}
\begin{minipage}[b]{0.45\textwidth}
\centering \includegraphics[trim = 15mm 17mm 17mm 15mm, clip, width=1\textwidth]{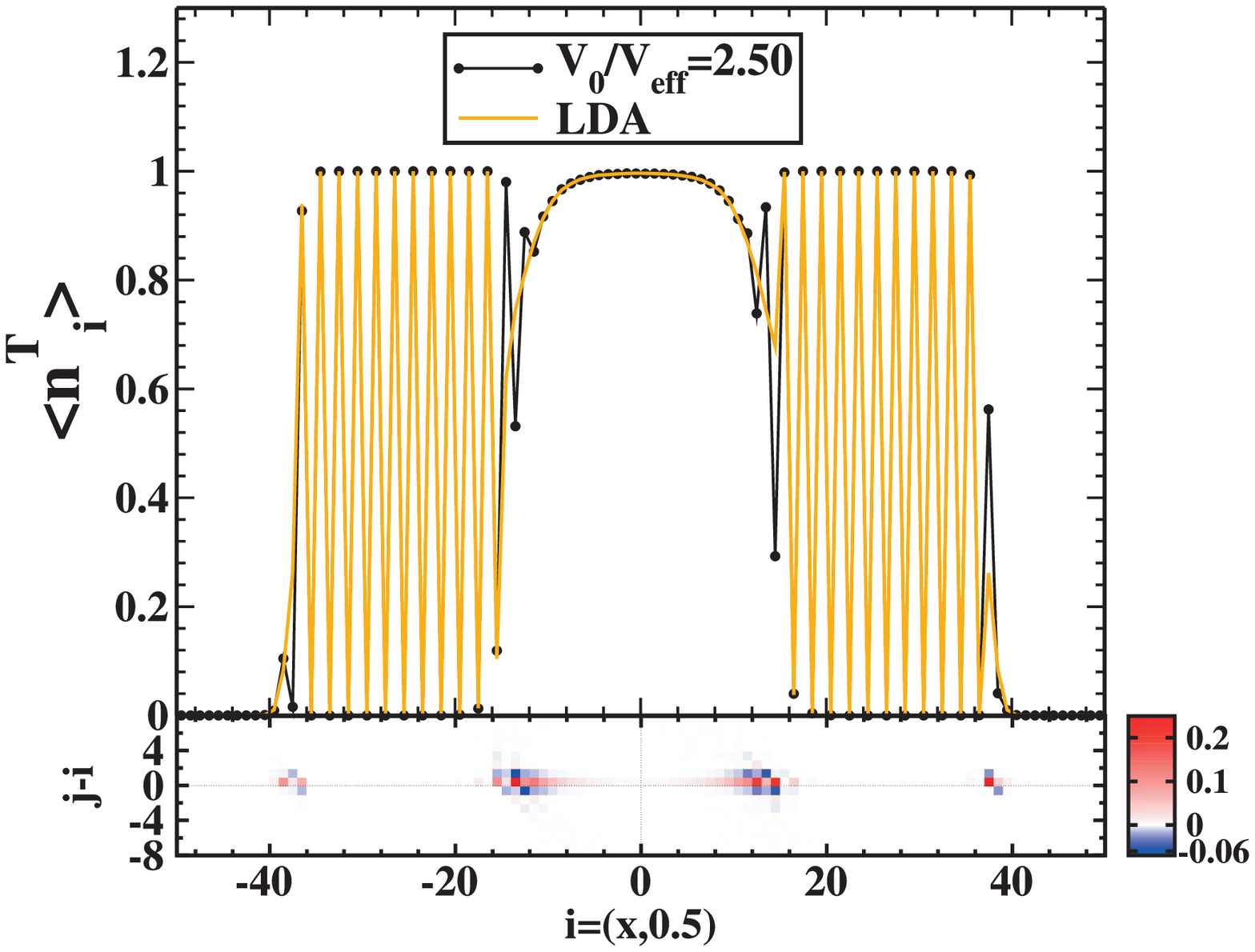}
\end{minipage}}\\
\caption{(Color online) Density profiles $\langle n_i^T\rangle$ along the cut $y=0.5$
obtained within our Monte Carlo simulation (black line with dots) for $k_BT_{\rm eff}/V_{\rm eff}=0.1$ 
and different values of the
harmonic confining potential $V_0/V_{\rm eff}=1$ (a), $V_0/V_{\rm eff}=1.95$ (b),  $V_0/V_{\rm eff}=2.05$ (c), and 
$V_0/V_{\rm eff}=2.50$ (d).
For comparison we also plot the LDA density profile 
(orange line). Below each panel we show a color-coded map plot of the correlation function 
$\Delta(i,j-i)$ for $i=(x,0.5)$ and $j=(x',0.5)$. (see detailed explanation in the text).}
\label{correlations}
\end{center}
\end{figure*}

\subsection{Trapped case}
Having studied the behavior of the homogeneous lattice system, we now consider the effect of superimposing a harmonic
trapping potential. 

It is easy to understand that the overall effect of the external confinement  is to compress the system and increase 
the density in the center at the expenses of the interparticle interaction.  Therefore one would expect that the CDW ground 
state found in the homogeneous case will eventually include a band insulating core for large enough values of the confining potential. 
The critical threshold for this phenomenon can be calculated analytically following the simple argument below. 

The energy of the outermost particle at the edge of the CDW phase is given by
\begin{equation}
\label{Energy_CDW}
E_{\rm CDW}=-\mu_{\rm eff}+V_0\frac{r_{\rm CDW}^2}{r_p^2}=-\mu_{\rm eff}+2V_0 \, ,
\end{equation}
where $r_{\rm CDW}=\sqrt{2}r_p$ is the CDW radius\cite{CDW_radius}, while the energy of this particle once moved to the trap
center (for simplicity we assume a central site at $(0,0)$ is given by 
\begin{equation}
\label{Energy_BI}
E_{\rm BI}=-\mu_{\rm eff}+4V_{\rm eff} \, .
\end{equation}
From here it directly follows that $E_{\rm CDW} < E_{\rm BI}$ for $V_0/V_{\rm eff} < 2$ and correspondingly it is energetically
more favorable to be in the CDW phase, while for $V_0/V_{\rm eff} > 2$ more particles will energetically 
prefer to move from the edges to the trap center and form a BI core in the center of the trap surrounded by a CDW ring. 

At zero temperature and for very high trapping potential, we can also expect the CDW-ring to disappear and the system to be in a
maximally packed state with radius $r_p$. The energy needed to move a single particle one step further out from the edge of the
maximally packed region can be easily evaluated. If this required energy is negative, the system will not be maximally packed in the
ground state. The energy of a particle at the edge of the maximally packed system sitting at radius $r_p$ is given by
\begin{equation}
E_p=\alpha V_{\rm eff}+V_0,
\end{equation}
with $\alpha$ being the average number of occupied nearest neighboring sites of the particle (a direct count yields $\alpha \approx 2.6$ for our system consisting of 2500 trions).
If this particle is moved one lattice step further out of the packed region, its energy is given by
\begin{equation}
E_o=(V_0/r_p^2)(r_p+a)^2=V_0\left(1+2a/r_p+a^2/r_p^2\right),
\end{equation}
Therefore, the condition for the CDW-ring to completely disappear is given by
\begin{align}
 \Delta E=E_p-E_o<0 \quad \Rightarrow \quad \frac{V_0}{V_{\rm eff}} > & \frac{\alpha}{2\frac{a}{r_p} +\frac{a^2}{r_p^2}}.
\end{align}
 
At zero temperature and $N=2500$ particles, $r_p/a=\sqrt{2500/\pi}\approx 28.2$ and the system would be maximally packed with no CDW-ring only for extremely large values of the trapping potential
\begin{equation}
\frac{V_0}{V_{\rm eff}} >36 \, ,
\end{equation} 
which are about $5$ times higher than the largest values considered in the numerics below.

In order to consider the properties of the system at finite temperature and for generic values of the Hamiltonian parameters, we address the system by using the Monte Carlo approach introduced above. 
We  always consider a two-dimensional square lattice with $M=100\times100$ sites, where the total number of trions is fixed to $N^T=N=2500$, while different values of the parameter $V_0$, which defines 
the strength of the harmonic confinement, and temperature $T$ are investigated.

As a first step we calculate the global CDW order parameter $C$ as a function of temperature $T$ for several values of $V_0$
(see Fig. \ref{CDW}). According to our calculations, at low temperature the CDW order parameter $C \not=0$, and therefore 
there is always a finite region where CDW order takes place for all the setups we considered ($V_0/V_{\rm eff}<6.5$).

As already mentioned in Sec. \ref{method}, we then further characterize the regions where CDW order takes place by investigating
the density profile $\langle n^T_i \rangle$ and the correlation function $\Delta(i,j)$. In order to represent the two-dimensional
system in our figures, we take a cross-section of the trap along the $y=0.5$ line, which is the closest cut to the center 
of the system. 
  
In Fig.~\ref{correlations} we fix the temperature as $k_BT/V_{\rm eff}=0.1$ and present results for different values
of the harmonic confining potential. For each setup we plot the density profile $\langle n^T_i \rangle$ along the cut $y=0.5$ 
as a function of the $x$-coordinate of the lattice site $i=(x,0.5)$. Below each panel we also show  the corresponding correlation
function $\Delta(i,j-i)$ in a color-coded map as a function of the distance $j-i=(x'-x,0)$ of the two points along the line $y=0.5$.

For comparison with the full Monte-Carlo results for the inhomogeneous system, we also plot the density profiles obtained 
within the local density approximation (LDA), which makes use of the Monte-Carlo results for the homogeneous system
with an effective local chemical potential $\mu_i=\mu_{\rm eff}-V_i$ for each lattice site along the trap. Due to the presence of two 
inequivalent sublattices in the CDW phase, the LDA density profile is drawn by considering pairs of 
neighboring lattice sites corresponding to different sublattices in the homogeneous case.

\begin{figure}[t]
\includegraphics[width=0.35\textwidth]{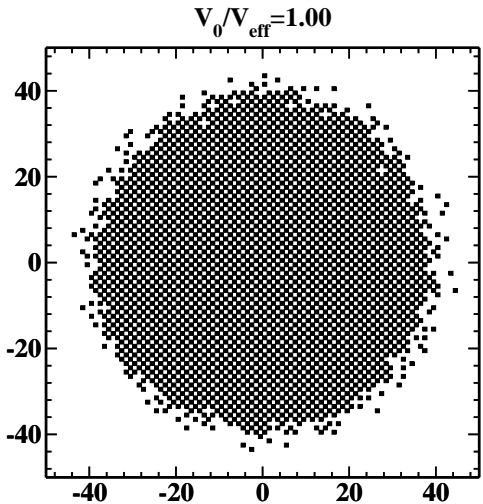}  
\vspace{-0.2cm}
\caption{Typical configuration of the trapped system for $V_0/V_{\rm eff}=1.00$ and $k_BT/V_{\rm eff}=0.1$, showing the presence
 of density fluctuations close to the edges of the CDW region.}
\label{snap_low}
\end{figure}

For weak confinement ($V_0/V_{\rm eff}=1$ in Fig. \ref{correlations}a), the density profile clearly shows staggered density
modulations in the trap center, centered around half-filling. These density modulations decay very fast at the border of the 
CDW region, where the density decays to zero within few lattice sites. In this case the LDA provides a fairly good description 
of the density profile inside the CDW region, while deviations can be observed close to the edges. A snapshot of a typical
configuration of the system for this setup is shown in Fig. \ref{snap_low}.

In order to better understand these deviations from the LDA predictions, one has to consider the correlation function plotted below
the density profile in Fig. \ref{correlations}, whose behavior provides information about the density fluctuations and the
correlation length. Indeed, for $i=j$, $\Delta$ provides the variance of the density distribution at the site $i$ along the
line $y=0.5$, therefore giving account of the number density fluctuations in the trap. At the temperature under investigation, the
thermal fluctuations in the bulk of the CDW regions are negligible, while sizable fluctuations can be observed at the borders of
the CDW regions, as evident also in Fig. \ref{snap_low}.

\begin{figure}[htb]
\includegraphics[width=0.35\textwidth]{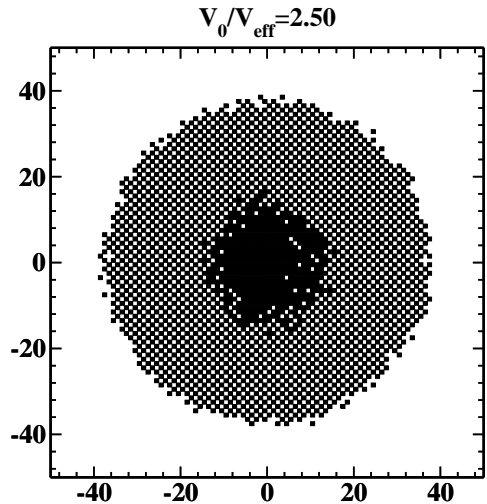}  
\vspace{-0.2cm}
\caption{Typical configuration of the trapped system for $V_0/V_{\rm eff}=2.50$ and $k_BT/V_{\rm eff}=0.1$. 
The presence of a band insulating core surrounded by a CDW ring is evident, with enhanced density fluctuations at the interface between the two regions.}
\label{snap_high}
\end{figure}

The density fluctuations show a staggered behavior in correspondence to the majority and minority sublattices of the 
disappearing CDW pattern. Only close to the borders of the CDW domain there is a sizable spread of the correlation function
with the distance $j-i$, where $\Delta$ also has a characteristic staggered behavior that vanishes within a few lattice sites.
The apparent left-right asymmetry in the correlation plots is due to our choice in the lattice position with respect to the 
center of the trap, since the first lattice sites around the center
are at $x=\pm 0.5$ and there is no central lattice site. At this point it is convenient to define the central average 
filling of the trap as 
\begin{equation}
n_{\rm center}^T=\frac{ \sum_{i=(\pm 0.5, \pm 0.5)} \langle n^T_i \rangle}{4},
\end{equation}   
where $n_{\rm center}^T=0.5$ for the setup considered in Fig.~\ref{correlations}a.

\begin{figure*}[htbp]
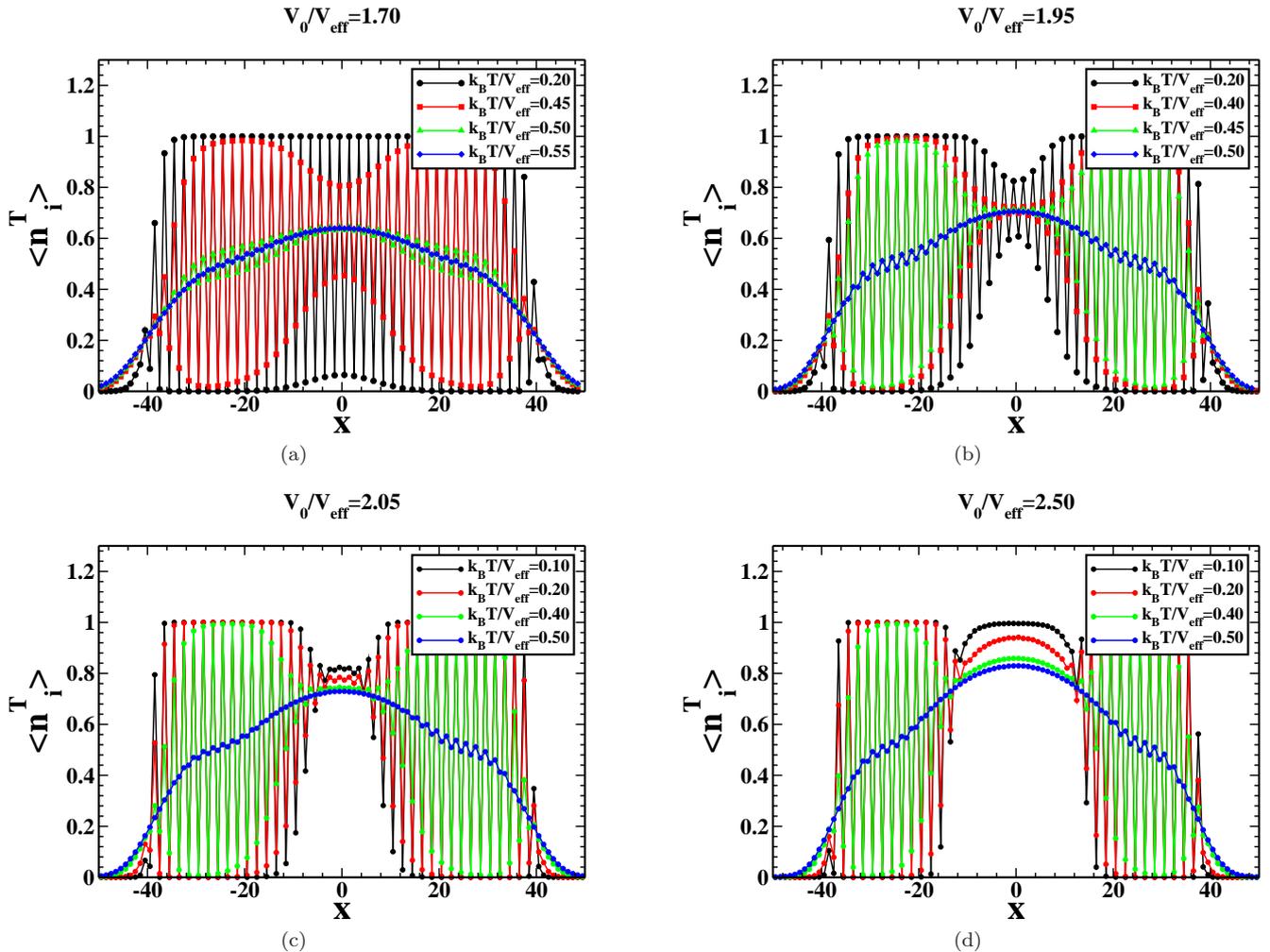

\begin{center}
\subfigure[]{
\begin{minipage}[b]{0.45\textwidth}
\centering \includegraphics[width=1\textwidth]{v1.70_diffT.eps}
\end{minipage}}
\hspace{1cm}
\subfigure[]{
\begin{minipage}[b]{0.45\textwidth}
\centering \includegraphics[width=1\textwidth]{v1.95_diffT.eps}
\end{minipage}}\\
\subfigure[]{
\begin{minipage}[b]{0.45\textwidth}
\centering \includegraphics[width=1\textwidth]{v2.05_diffT.eps}
\end{minipage}}
\hspace{1cm}
\subfigure[]{
\begin{minipage}[b]{0.45\textwidth}
\centering \includegraphics[width=1\textwidth]{v2.50_diffT.eps}
\end{minipage}}\\
\caption{(Color online) Monte-Carlo density profile $\langle n_i^T\rangle$ for four different values 
of the harmonic confining potential  $V_0/V_{\rm eff}=1.70$ (a), $V_0/V_{\rm eff}=1.95$ (b),
$V_0/V_{\rm eff}=2.05$ (c), and $V_0/V_{\rm eff}=2.50$ (d), and different temperatures $k_BT_{\rm eff}/V_{\rm eff}$.}
\label{dens_vs_T}
\end{center}
\end{figure*}

With the increase of the trapping potential strength, shown in  Fig.~\ref{correlations}b,~\ref{correlations}c and ~\ref{correlations}d, there is a gradual suppression 
of the density modulations in the trap center and the CDW region smoothly migrates away from the trap center evolving into a 
CDW ring for large values of the trapping potential. It is worth noting, e.g. in Fig.~\ref{correlations}b, that 
the local density in the central region of the trap fluctuates around an average density which is significantly higher than half-filling
and approaches $n_{\rm center}^T \approx 0.75$ for the setup in Fig.~\ref{correlations}b. For increasing trapping potential
the density modulations eventually disappear in the center (see Fig.~\ref{correlations}c), leading to an unordered trionic 
phase whose density increases with $V_0$.

As evident in Figs.~\ref{correlations}b and \ref{correlations}c, the LDA approach is unable to properly describe 
the density profile in this regime of parameters, since within LDA the density modulations in the center disappear 
abruptly for much smaller values of the confinement. The presence of wider stability regions for CDW
ordering within the full Monte-Carlo calculation has to be explained through a proximity effect: CDW order is induced 
in the trap center from the surrounding ring in the case $n_{\rm center}^T > 0.5$, where for the homogeneous case this value of the density is too high to stabilize the 
CDW phase. We recall that in the homogeneous case the CDW order is always commensurate 
in our findings and $\langle n^T \rangle=0.5$. This proximity effect in the trap can be clearly understood by looking at the correlation functions,
which show the existence of non-zero correlations between different sites in the central region of the trap;
a feature that is omitted in the LDA description. A similar effect has also been observed for antiferromagnetic order in a harmonic trap\cite{RDMFT}.    

Finally, for strong confinement ($V_0/V_{\rm eff}=2.5$ in Fig.~\ref{correlations}d), we found that the trions are in a close packed
arrangement in the central region and therefore form a band insulating (BI) core in the trap center, surrounded by a CDW ring,
as evident also in Fig. \ref{snap_high}.
The area of this ring decreases for increasing trapping potential, but as previously mentioned it never vanishes completely for the values of the confinement potential 
we considered in our calculations ($V_0/V_{\rm eff}<6.5$).

The evolution of the CDW regions at fixed values of $V_0/V_{\rm eff} > 1.5$ and for increasing temperature $T$ is displayed in
Fig.~\ref{dens_vs_T} through the density profile. The CDW ordered region in the center melts at a temperature which slightly 
decreases as $V_0$ is increased from $V_0=1.5$ in Fig.~\ref{dens_vs_T}a to $V_0=1.95$ in Fig.~\ref{dens_vs_T}b, while the CDW order in
the surrounding ring is clearly much more robust against thermal fluctuations. Correspondingly the asymptotic value of the density
in the trapping center $n_{\rm center}^T \geq 0.5$ at the melting point for the CDW core increases as a function of $V_0$. 
For a strongly confining trap ($V_0=2.5$ in Fig.~\ref{dens_vs_T}d), the CDW order is never present in the trap center, however, 
increasing the temperature causes also the BI core to melt. Further increase of the temperature has a limited
effect on the remaining unordered trionic phase.

\begin{figure}[hbpt]
\includegraphics[width=0.45\textwidth]{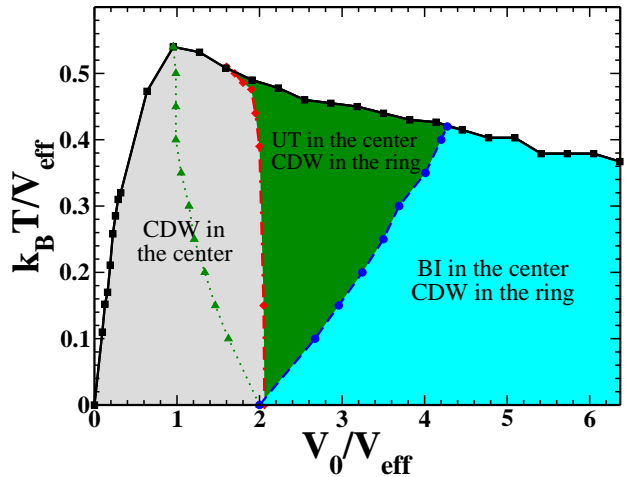}  
\vspace{-0.2cm}
\caption{(Color online) Phase diagram of the two-dimensional trionic system in the presence of harmonic confinement. 
The black solid line marks the critical temperature $T_c$ for the complete disappearance of CDW order in the system.
The gray area below this line and left of the red dashed-dotted line corresponds to CDW order in the trap center surrounded by
unordered trions (UT). Within this region the dotted green line marks the transition between a CDW with $n^T_{\rm center}=0.5$ and
$n^T_{\rm center} > 0.5$. The green area below the black solid line and between the red dashed-dotted and blue dashed lines
corresponds to CDW order in a ring, while inside and outside of the ring the trions are in an unordered trionic phase. The
unordered trionic region at the trap center evolves into a band insulator (BI) within the cyan area, underneath the
black solid line and right of the blue dashed line. At the interface between the BI and CDW regions as well as outside of CDW ring
the trions are in an unordered phase. Finally, above the black solid line the whole system is in an unordered trionic phase
since the CDW order disappears due to thermal fluctuations.}
\label{phase_diagram}
\end{figure}

Our results for the trapped inhomogeneous system are summarized in the phase diagram in Fig.~\ref{phase_diagram}. 
The black solid line in Fig.~\ref{phase_diagram} always marks the critical temperature $T_c$ for complete disappearance of 
CDW order in the system, irrespective of the position of the CDW domain. Depending on the harmonic confinement $V_0/V_{\rm eff}$ 
and temperature $k_B T/V_{\rm eff}$ we get four different scenarios for the spatial arrangement of trions:
\begin{itemize}
 \item
A large region of CDW order in the trap center surrounded by unordered trions for small values of the confinement strength $V_0$ 
and $T < T_c$ (gray area, below black solid line and left from red dashed-dotted line in Fig.\ref{phase_diagram}). We have 
estimated the entropy per particle $S(T)=\int_0^T C(T')/T' dT'$, where  $C(T')$ is the specific heat, at the maximum critical temperature
$K_B T_c/V_{\rm eff} = 0.54$ to be around $1.4k_B$.
Within this region we can further distinguish between two different cases:
(i) For low $T$ and small $V_0$, the average central filling $n^T_{\rm center}=0.5$ (left from green dotted line) and (ii) for increasing
$T$ and $V_0$ the average central filling 
is  $n^T_{\rm center} > 0.5$ (right from green dotted line). At low but finite temperature the CDW core 
in the trap center disappears for $V_0/V_{eff}=2$ and $n^T_{\rm center} = 0.75$ at the transition point. 
\item
For intermediate values of the confinement strength there is CDW order in a ring around the trap center, while in the central region 
and outside of the CDW ring the trions are in an unordered phase (green area below black solid line and between red dashed-dotted and blue
dashed lines).
\item
For large values of the confinement strength and low temperatures there is a BI phase of trions in the center, while CDW order takes 
place in the ring surrounding it (cyan area, below black solid line and right from blue dashed line).
At the interface between the BI and the CDW ring as well as outside of the CDW ring, the trions are in an unordered phase.
\item
The complete system is in an unordered trionic phase for high temperatures (white area above black solid line).
\end{itemize}

\begin{figure}[hbpt]
\includegraphics[width=0.45\textwidth]{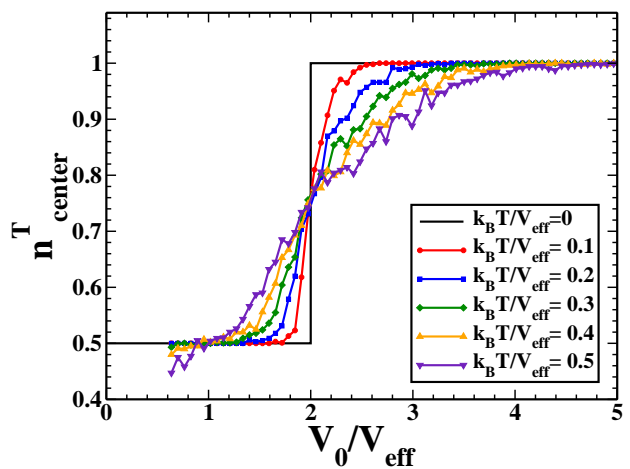}  
\vspace{-0.2cm}
\caption{(Color online) Average central filling $n^T_{\rm center}$ as a function of the harmonic confinement potential $V_0/V_{\rm eff}$
for different temperatures.}
\label{center_filling}
\end{figure}

We note that, as one can see in Fig. \ref{center_filling}, for $T>0$ the average central 
filling $n^T_{\rm center}$ is a continuous function of the trapping potential and increases for increasing $V_0$,
while at $T=0$ the occupation changes discontinuously from $n^T_{\rm center}=0.5$ to $n^T_{\rm center}=1$. 
In the latter case, the system stays in a CDW ground state until it becomes energetically favorable to move trions with 
large potential energy from outside the CDW domain to the central region at the expense of the interaction energy
according to the simple analytical argument given in the beginning of this subsection.

\section{CONCLUSION}\label{conclusion}

In this paper we have investigated strongly attractive three-component Fermi gases loaded into an optical lattice and have explicitly
taken the effect of a harmonic confinement into account. We considered the effective strong-coupling Hamiltonian for the trionic phase derived by us in Ref.~[\onlinecite{NJP}] 
and showed that, at the leading order in $J/|U|$, the trionic hopping can be safely neglected. Hence the effective Hamiltonian  describes asymptotically immobile trions with nearest-neighbor 
interaction. Since this model has a structure analogous to the antiferromagnetic Ising model, we performed classical Monte-Carlo simulations to study the spatial arrangement of the trionic 
particles in this limit.  

First we considered a homogeneous system without harmonic confinement. In this case the Hamiltonian is equivalent to an
antiferromagnetic Ising model in a uniform magnetic field. The results of our approach were found to be in very good agreement with
previous results for the Ising spin model. They show that the trions are arranged in a staggered density wave configuration at
half-filling while they are unordered for an incommensurate density. When taken into account the subleading kinetic energy, 
the  degeneracy of this unordered phase is likely to be lifted, letting the system evolve into a homogeneous metallic phase, as expected 
according to the available literature on the spinless fermion model at strong-coupling\cite{gubernatis,vlaming}. We determined the critical 
temperature for the disappearance of the CDW order, which is analogous to the critical temperature of the equivalent spin model. 

In the presence of a harmonic trap we found several possible scenarios for the spatial arrangement of trions, namely coexistence of CDW domains with unordered trions and band insulating regions, 
depending on the strength of the confining potential
and temperature. The CDW region moves from the center of the trap to a ring for increasing trapping potential. We found that
the staggered density order is also induced, due to a proximity effect, in regions of the trap where the average density
is not commensurate, in close analogy to the behavior at the edges of the the antiferromagnetic domain in trapped two-component Fermi mixtures \cite{RDMFT}. The inclusion of spatial correlations 
is crucial for a proper description of this feature, which is indeed missing in a LDA description of the system.

An important question arising from this investigation is how the behavior of the trions is modified by the inclusion of a finite
hopping, considering that already the homogeneous case gives rise to a very rich phase diagram\cite{gubernatis,vlaming}.
A mapping onto an equivalent spin Hamiltonian would lead, however, to a strongly anisotropic XXZ model\cite{capone_dao}, and would therefore require an extension from a classical to a quantum Monte Carlo 
approach. We postpone this interesting question to a future work. 

\section*{Acknowledgment}

We thank Selim Jochim for insightful discussions about three-component Fermi gases. 
AP thanks Massimo Capone for providing additional references,  European Research Council under 
FP7/ERC Starting Independent Research Grant “SUPERBAD” (Grant Agreement n. 240524) for financial support, 
Claudio Verdozzi and the Division of Mathematical Physics at Lund University (Sweden) for their hospitality during the completion of this work. SB thanks Daniel Cocks for useful suggestions. 
This work was supported by the German Science Foundation DFG via Sonderforschungsbereich SFB-TRR~49.


\begin{thebibliography}{99}
\bibitem{virtual} The scaling of the effective hopping parameter is easily understood since moving a trion to a neighboring site
involve three different virtual jumps.
\bibitem{CDW_radius} 
The area occupied by the system in CDW order ($A_{\rm CDW}=\pi r_{\rm CDW}^2$) is twice larger then area for maximally packed system 
($A_{\rm p}=\pi r_{\rm p}^2$), therefore  $r_{\rm CDW}=\sqrt{2}r_p$.
\end{thebibliography}
\end{document}